\documentclass[aps,prl,twocolumn,groupedaddress]{revtex4}

\usepackage[dvipdfmx]{graphicx}
\usepackage{float}
\begin{document}
%\preprint{}

\title{First-principles study of spin texture and Fermi lines in Bi(111) multi-layer nanofilm}


\author{Hiroki Kotaka$^{1,2,3,4}$,  Fumiyuki Ishii$^{2}$ and Mineo Saito$^{2}$}
%\email[]{kotaka@esicb.kyoto-u.ac.jp}
%\thanks{}
\affiliation{ 
\small $^{1}$ Graduate School of Natural Science and Technology, Kanazawa University Kanazawa 920-1192, Japan\\
\small $^{2}$ Institute of Science and Engineering, Kanazawa University, Kanazawa 920-1192, Japan\\
\small $^{3}$ ISIR-SANKEN, Osaka University, 8-1 Mihogaoka, Ibaraki, Osaka, 567-0047, Japan\\
\small $^{4}$ ESICB, Kyoto University, 1−30, Goryoohara, Nishikyo-ku, Kyoto 615-8510, Japan\\
}


\date{\today}
\newpage
\begin{abstract}
We have performed a fully relativistic first-principles density functional calculation 
examining the surface state of bismuth (Bi) (111) multi-layer nanofilm, 
with up to 20 Bi bilayers, and investigated the Rashba effect and spin texture on the Bi surfaces. 
We have revealed a giant out-of-plane spin states on the Fermi lines, 
% based on the Fermi linesでよいのか疑問？
and the maximum value of the out-of-plane spin component being approximately 40$\%$ of the magnitude of the total spin. 
We have also evaluated the Rashba parameter $\alpha_R \simeq 1.9 {\rm eV}\cdot$\AA~
using the surface state bands which is buried in the bulk state, at -0.32 eV below the Fermi energy. 
%もっと追加の文章を増やせといわれた（ジャーナルに合わせて、広い範囲の興味の為の更なる文章の追加が必要）
\end{abstract}

%\keywords{}

\maketitle

\section{Introduction}
Understanding the transport properties of bismuth(Bi) has long been a major 
aim in the field of condensed matter physics. 
The anomalous transport properties occur in this material, 
originating from its semimetallic electronic structures\cite{Issi1979}, 
in which electron and hole carriers coexist.
In addition, the Bi atom has a strong spin-orbit interaction(SOI) originating from its high atomic number(83).
Therefore, it is expected that various relativistic effects apply, such as the Rashba effect\cite{RashbaEffect_SovPhys1960}, the quantum spin Hall effect\cite{Murakami_ZBNR_prl2006}, and conversion between the spin and charge currents\cite{SpinChargeConv_ncomm_2013}.
%kotaka

%第2段落
Bi films are expected to be suitable for device application\cite{Nagao_prl2004} 
as it has been shown that flat thin films can be grown on a Si (111) surface.  
In this case, the surface state differs strongly from the bulk state; 
that is the surface density of states is significantly larger than that of the bulk states\cite{Hirahara_prl2006}.
As a result of the strong SOI within the electronic structures, spintronic applications are expected.

%第3段落
Recent angle-resolved photo-emission spectroscopy (ARPES) experiments \cite{Takayama_Pz_prl2011} 
have revealed the novel spin textures of the surface states of Bi thin films. 
The conventional Rashba (in-plane) spin texture and a giant out-of-spin component are detected. 
The spin texture has been analyzed using the tight-binding model\cite{BT-Bifilm}; 
quantitative analysis of the spin textures based on the first-principles calculations has not been conducted. 

In this paper, we conduct a theoretical quantitative investigation of the surface states of Bi (111) multi-layer nanofilm based on the Fermi lines. 
Using fully relativistic first-principles density functional calculations, 
we perform quantitative analysis for the surface localized spin vortexes on the Fermi lines of the Bi thin film.
Our calculation supports the experimentally observed large out-of-plane spin component.
Further, we find that the Rashba parameter $\alpha_R$ of these bands is large (approximately 1.9 $\pm$ 0.1 eV). \\

\section{theoretical method}

Using the OpenMX code\cite{OpenMX}, 
we perform fully relativistic first-principles calculations based on density functional theory (DFT) within the generalized gradient approximation (GGA)\cite{PhysRevLett.77.3865_PBE}.
In our study, wavefunctions are expressed as a linear combination of multiple pseudo atomic orbitals (LCPAO) generated by a confinement scheme\cite{PhysRevB.72.045121_LCPAO, PhysRevB.67.155108_LCPAO, PhysRevB.69.195113_LCPAO}.

We deduce the spin polarization in the reciprocal lattice vector \textbf{k} from the spin density matrix. 
The spin density matrices $P_{\sigma\sigma'}(\textbf{k},\mu)$  are calculated using the spinor Bloch wavefunction, 
the component of which is given by $\psi_{\mu}^{\sigma}(\textbf{r},\textbf{k})$, 
which is obtained from the OpenMX calculations, where $\sigma$ is the spin index ($\uparrow$ or $\downarrow$) and $\mu$ is the band index. 
Thus, 
\begin{eqnarray}
P_{\mu\sigma\sigma'}(\textbf{k}) & = & \int \psi_{\mu}^{\sigma}(\textbf{r},\textbf{k})\psi_{\mu}^{\sigma'}(\textbf{r},\textbf{k})d\textbf{r} \label{eq:MulP}\\
 & = & \sum_{n}\sum_{i,j} \left[ c_{\sigma\mu i}^{\ast}c_{\sigma'\mu j}^{ }S_{ij}\right] e^{i\textbf{R}_n\cdot\textbf{k}}, \label{eq:Pcal}
\end{eqnarray}
where $S_{ij}$ is the overlap integral of the $i$-th and $j$-th localized orbitals and $c_{\sigma \mu j}$ is expansion coefficient. 
$\textbf{R}_n$ is the $n$-th lattice vector.
We evaluated spin vector component ($P_{x}$, $P_{y}$, $P_{z}$) from density matrices.\\

We perform first-principles calculations of 20-bilayer Bi (111) film (Fig. 1). 
The length of the in-plane unit vector $a$ is 4.53 \AA, 
and the thickness of the film is 76.5 \AA. 
Further, each bilayer forms a honeycomb lattice and has 1.6 \AA-buckling. 
\begin{figure}[t]
\begin{center}
\includegraphics[width=7cm]{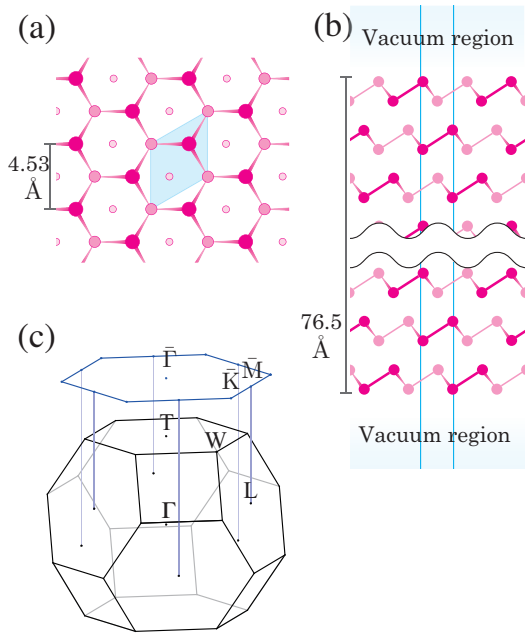}
\end{center}
\caption{(a)Top view and (b) side view of 20-bilayer Bi (111) thin film. (c) Brillouin zone of Bi. Black line shows Brillouin zone of bulk structure, and the blue line shows Brillouin zone of (111) surface structure.}
\label{f1}
\end{figure}

We confirm that the Fermi lines are insensitive to the thickness when more than 13 bilayer films are used. \\

In the 20-bilayer slab calculation, the interaction between the two surfaces is expected to be very weak and can be neglected.
As the slab has inversion symmetry, the Kramers degenerate bands appear. %Kramer’s degeneracy bandsを推奨されたが、このまま
Because of the weak interaction between two surfaces, unitary transformation of the two bands generates to wavefunctions that are localized at each surface. 
By analyzing  the wavefunctions, we can obtain information on the surface states.

\section{Band structure and surface states}

\begin{figure}[t]
\begin{center}
\includegraphics[width=8.5cm]{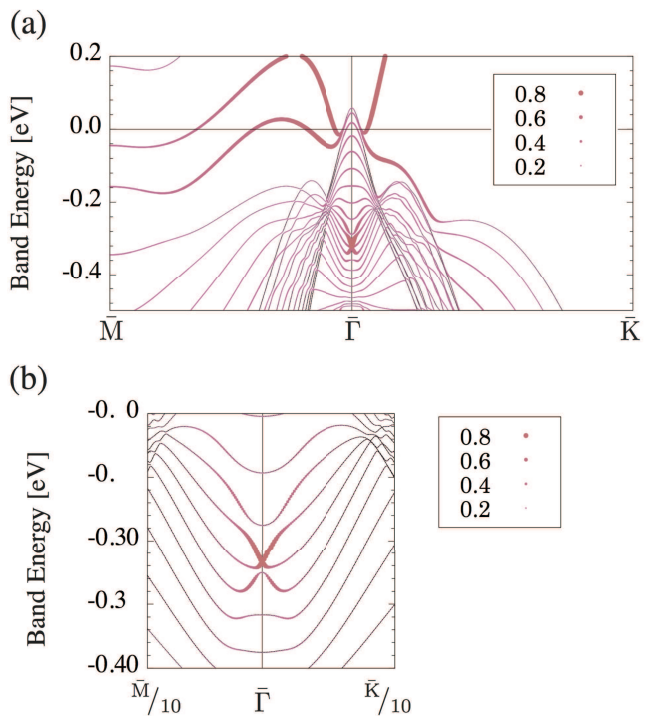}
\end{center}
\caption{(a)Surface-projected band dispersion (b) enlarged figure near $\bar{\Gamma}$ of 20-bilayer Bi (111) thin film. The circle radius and color show the surface-localization ratio projected surface-1bilayer. }
\label{f2}
\end{figure}

Figure \ref{f2}(a) shows the band structure of the 20-bilayer Bi (111) thin film. 
The surface localization ratio is represented by the circle radius proportional to the number of electrons within the surface-1bilayer. 
Near the Fermi level $E_{\rm {F}}$ along the $\bar{\rm M}$-$\bar{\Gamma}$ line, 
two surface-localized bands appear. 
The surface bands are connected to the bulk-like bands around the $\bar{\Gamma}$ point near the $E_{\rm {F}}$, 
at which states the bands have small amplitude on the surface layer. 
The bulk-like bands originate from the hole pocket at the $T$ point in the bulk structure Bi%\cite{Freeman}. 

On the other hand,  as shown in fig. \ref{f2}(b),
the surface localized states at the $\bar{\Gamma}$ point are buried in the bulk-like state in the region of $E$\ -\ $E_{\rm {F}}$= -0.32 eV, where $E$ is band energy. 
More than 50\% of these states are localized in the surface 1-bilayer. 
These bands cross at the $\bar{\Gamma}$ point, and the energy contours form circular shape.
In addition, the spin texture indicated by these bands corresponds to a typical in-plane spin vortex. 
These features appear to conform to a typical Rashba band.
We estimate the Rashba parameter $\alpha_R$ from the gradient of these bands.
The estimated $\alpha_R$ is 1.9 $\pm$ 0.1 eV$\cdot$\AA, which is very large and is of the same order as the value obtained for other bismuth compounds (e.g., BiTeI has $\alpha_R$ = 3.8eV$\cdot$\AA  \ \cite{Ishizaka_BiTeI_NatMat2011}).

\begin{figure}[t]
\begin{center}
\includegraphics[width=8.5cm]{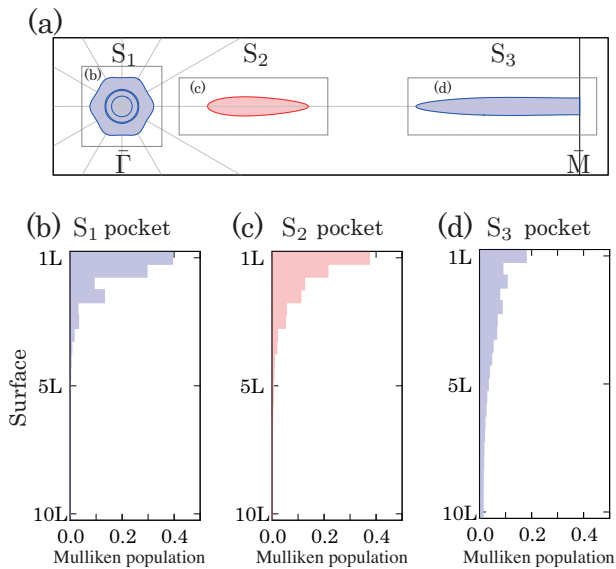}
\end{center}
\caption{(a) Fermi lines of 20-bilayer Bi (111) thin film. (b)-(d) show the electron density of each layer at S$_1$-S$_3$ pocket.
One graph bar means one atom, and  two graph bar means 1-bilayer.}
\label{eleden}
\end{figure}

\section{Fermi lines and surface localization}

Figure \ref{eleden}(a) shows the Fermi lines of a freestanding 20-bilayer Bi (111) film. 
There are two electron pockets (S$_1$ and S$_3$) and one hole pocket (S$_2$).
S$_1$, S$_2$, and S$_3$ have hexagonal, elliptical, and needle-like shapes, respectively.
These features are consistent with those observed in previous studies based on ARPES experiment\cite{Hirahara_prl2006, Takayama_Pz_prl2011}.

To perform quantitative analysis of the Fermi lines, we estimate the area of each pocket and compare the results with the ARPES-derived experimental values\cite{Takayama_Pz_M_nl2012}. 
The calculated areas of the S$_1$, S$_2$ and S$_3$ pockets are 8.7 $\times$ 10$^{-3}$[\AA $^{-2}$], 4.4 $\times$ 10$^{-3}$[\AA $^{-2}$] and 7.7 $\times$ 10$^{-3}$[\AA $^{-2}$], respectively. 
The area of each pocket are close to those observed in the ARPES experiments\cite{Takayama_Pz_M_nl2012}. 

Next, in order to investigate the surface localization states, we calculate the Mulliken population from the LCPAO parameter.
In the S$_{1}$ and S$_{2}$ pockets, more than 60\% of the wave functions components are in the surface 1-bilayer and more than 90\% of the components are in the surface 2-bilayer (Fig. \ref{eleden}(b)- (c)).
The electron density derived from the S$_{3}$ pocket is more delocalized than those for the S$_{1}$ and S$_{2}$ pockets, and only 30\% of the wavefunction components are in the surface 1-bilayer(Fig. \ref{eleden}(d)).
The surface localization of each pocket corresponds to the intensities obtained from the ARPES experiments; 
therefore S$_{1}$ and S$_{2}$ pockets are expected to have strong intensity, while the S$_{3}$ pockets are expected to have weak intensity. 
This is consistent with the ARPES experiments\cite{Takayama_Pz_M_nl2012}.
The difference in localization for each of the Fermi lines originates from the direct band gap of bulk Bi. 
The direct energy gaps in the bulk Bi are 0.26 eV and 0.11eV at the $T$ and $L$ points (Fig.1 (c)), respectively, 
where the $T$ and the $L$ points of the bulk Bi correspond to the $\bar{\Gamma}$ and $\bar{\rm M}$ points of the Bi film, respectively.
The small direct energy gap may cause the surface delocalization at the $\bar{\rm M}$ point. 

\begin{figure}[t]
\begin{center}
\includegraphics[width=8cm]{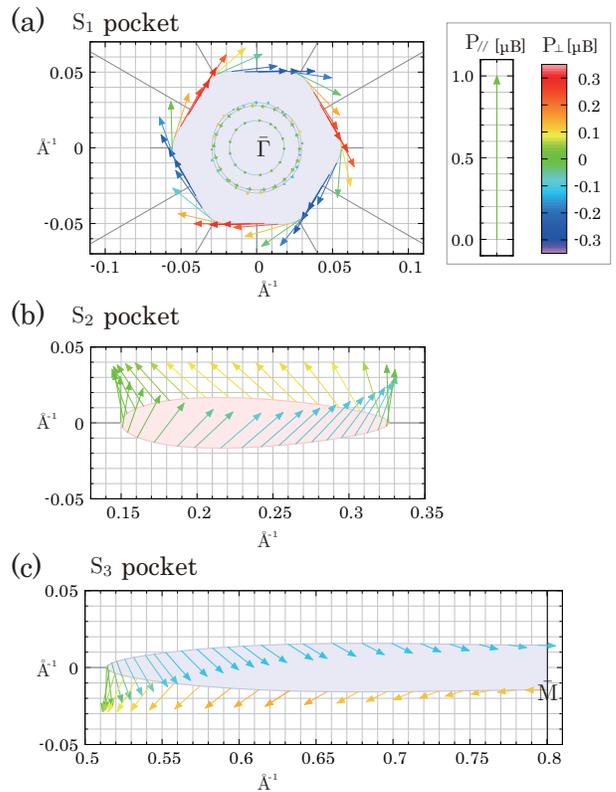}%FS_spin2.eps
\end{center}
\caption{Surface-localized spin textures of (a) S$_1$ pocket (b) S$_2$ pocket, and S$_3$ pocket. The vector length of each arrow express the magnitude of in-plane spin component, and the color of each arrow express out-of-plane spin component.}
\label{f4}
\end{figure}

\section{Surface-projected spin texture in momentum space}
Next, we discuss whether the spin textures on the Fermi lines are typical in-plane Rashba-type spin or non-Rashba-type spin vortices including a giant out-of-plane spin component. 

Figures \ref{f4}(a) - (c) show the Fermi lines and surface-localized spin texture (in-plane spin component P$_{\// \! \//}$, out-of-plane spin component $P_{\perp}$) projected onto surface 1-bilayer.
The vector length shows the magnitude of $P_{\// \! \//}$, whereas that of $P_{\perp}$ is expressed by the color bar.
In S$_{1}$ pocket, $P_{\// \! \//}$ has clockwise vortex and $P_{\perp}$ has $2\pi/3$ rotational periodicity.
%The S$_{1}$ pocket has an in-plane clockwise spin vortex and the out-of-plane spin component P$_{\perp}^{k}$ which has $2\pi/3$ rotational periodicity.
The magnitude of $P_{\perp}^{k}$ has the maximum value at $\bar{\Gamma}-\bar{\rm K}$ line and minimum value 0 at the $\bar{\Gamma}-\bar{\rm M}$ line.
%The periodicity of P$_{\perp}^{k}$ (max:$\bar{\Gamma}-\bar{\rm K}$ and 0:$\bar{\Gamma}-\bar{\rm M}$) are shown 
%and we compare the in-plane spin with out-of-plane spin P$_{\perp MAX}^{k}$.
The maximum magnitude of the $P_{\perp}^{k}$ is 40\% of the spin vector.
These features are in reasonably good agreement with the result of ARPES experiment\cite{Takayama_Pz_prl2011} 
and can be explained via group theoretical analysis.\\

The obtained spin textures in momentum space depend on the symmetry group.
The first-order spin Hamiltonian over the wave vector is expressed as 
$H = \sum_{i,j} \alpha_{ij} k_i \sigma_j\label{eq:HSO_1}$, 
where $k_i$ is the $i$-th component  of the wave vector and $\sigma_j$ is the Pauli-matrix. 
$i$ ($j$) runs over $x$ and $y$ ($x$, $y$, and $z$). 
The spin Hamiltonian satisfies the following expression via effective symmetric operation :$R^{-1} H R = H$.
The $\bar{\Gamma}$ point has $C_{3v}$ symmetry. (if we consider the two surfaces, the symmetry is $D_{3d}$.)
The in-plane spin texture in the Bi surface can be explained by the first-order term of the spin Hamiltonian, $H_{\Gamma}^1 = \alpha_R(k_x\sigma_y-k_y\sigma_x)$\cite{Oguchi_kp_2009}. 
On the other hand, the out-of-plane spin component is not explained by the first-order term, 
but is instead explained by the third-order term of Hamiltonian, expressed as $H_{\Gamma}^3 =\alpha_{31}k^2(k_x \sigma_y - k_y \sigma_x)+\alpha_{32}(k_y^3 - 3k_x^2k_y)\sigma_z$\cite{Vajna_RB3rd_prb2012}.
The out-of-plane spin components on the $\bar{\Gamma}\bar{\rm M}$ line are zero, because it is on the $k_x$-$k_z$ plane($k_y$=0).  
In the Bi surface, third-order is very important, and this feature is similar to the result of a our previous study on the Bi 1-bilayer\cite{kotaka_jjap_2012}.

The first-order term of the spin Hamiltonian on the $\bar{\Gamma}\bar{\rm M}$ line is expressed as $H_{\bar{\Gamma}\bar{\rm M}}=\alpha_1 k_x \sigma_y + \alpha_2 k_y \sigma_x + \alpha_3 k_y \sigma_z$. 
Therefore, the S$_2$ and S$_3$ pockets are consistent with this spin Hamiltonian.
The origin of out-of-plane spin component is different from S$_1$ pocket.

More detail information of spin textures is provided in supplementary\cite{suppl}.\\

\section{Summary}
We performed fully-relativistic first-principles density functional calculations of the surface state of the Bi (111) multi-layer nanofilm.
We investigated the surface-localized spin texture of the 20-bilayer Bi nanofilm, 
and we focused on the giant out-of-plane spin.% as found in the Fermi line.  % as indicated by
The maximum magnitude of the P$_{\perp}^{k}$ is 40\% of the spin vector.\\
In addition, we revealed the surface-localized typical Rashba spin-split state, 
which is buried in the bulk states.
The Rashba parameter $\alpha_R \simeq 1.9 \pm 0.1{\rm eV}\cdot$\AA \ was 
evaluated using these surface state bands.

When this work was almost completed, 
a recent paper \cite{BT-Bifilm} regarding the tight-binding calculation of spin split states in Bi thin film come to our attention. 
The out-of-plane spin and the differences in the localization of the Fermi lines reported in that study are qualitatively consistent with our first-principles results. \\

\section{acknowledgments}
The author would like to thank T. Oguchi, A. Takayama, T. Okuda and T. Aono for invaluable discussions.
Part of this research has been funded by the MEXT HPCI Strategic Program. This work was partly supported by Grants-in-Aid for Scientific Research (Nos. 25390008, 25790007, 26108708 and 15H01015) from the JSPS and by the RISS project in the IT program of MEXT. One of the authors (H.K.) thanks JSPS for the financial support (No. 22-2329). The computations in this research were performed using the supercomputers at the ISSP, University of Tokyo, the RCCS, Okazaki National Institute, and the IMR, Tohoku University.\\

%\bibliography{Bi_surf}
%\bibliographystyle{h-physrev3}
\bibliographystyle{apsrev}
%\bibliographystyle{PhysRevB}

%\end{document}
\newpage 
\widetext
%\\*
\begin{center}
	\textbf{\LARGE  Supplemental Materials}
\end{center}
\setcounter{figure}{0}
\setcounter{page}{1}
\makeatletter
\renewcommand{\theequation}{S\arabic{equation}}
\renewcommand{\thefigure}{S\arabic{figure}}
\renewcommand{\bibnumfmt}[1]{[S#1]}
%\newcommand{\citenumfont}[1]{S#1}

%\newpage
\begin{figure}[H]
\begin{center}
\includegraphics[width=15cm]{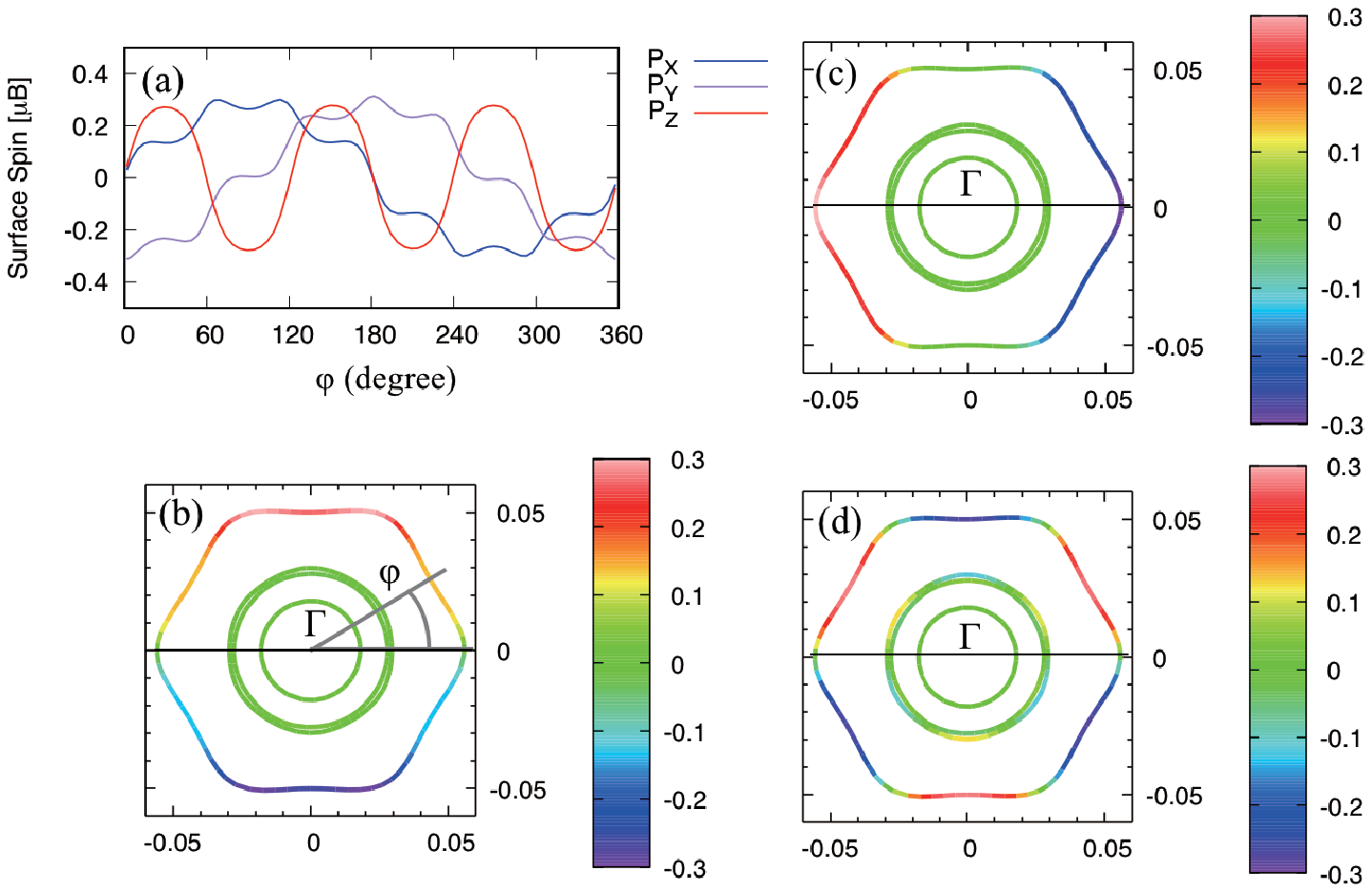}
\end{center}
\caption{Calculated spin texture of S$_1$ pocket. 
(a) Angle-dependet spin components $P_x$, $P_y$, and $P_z$, 
(b) Fermi line with $P_x$ component,
(c) Fermi line with $P_y$ component,
and (d) Fermi line with $P_z$ component.
}
\label{s1}
\end{figure}

\begin{figure}[H]
\begin{center}
\includegraphics[scale=1.0,angle=0]{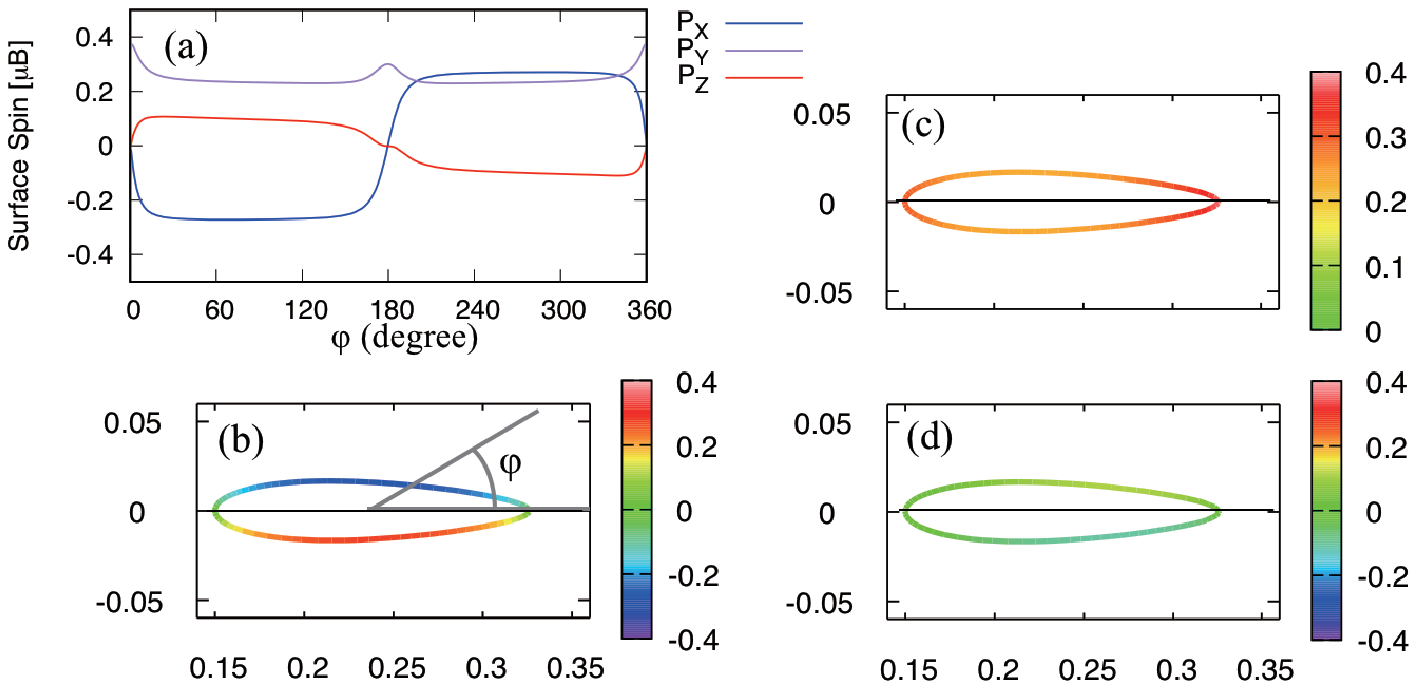}
\end{center}
\caption{Calculated spin texture of S$_2$ pocket. 
(a) Angle-dependet spin components $P_x$, $P_y$, and $P_z$, 
(b) Fermi line with $P_x$ component,
(c) Fermi line with $P_y$ component,
and (d) Fermi line with $P_z$ component.
}
\label{s2}
\end{figure}

\begin{figure}[H]
\begin{center}
\includegraphics[width=10cm]{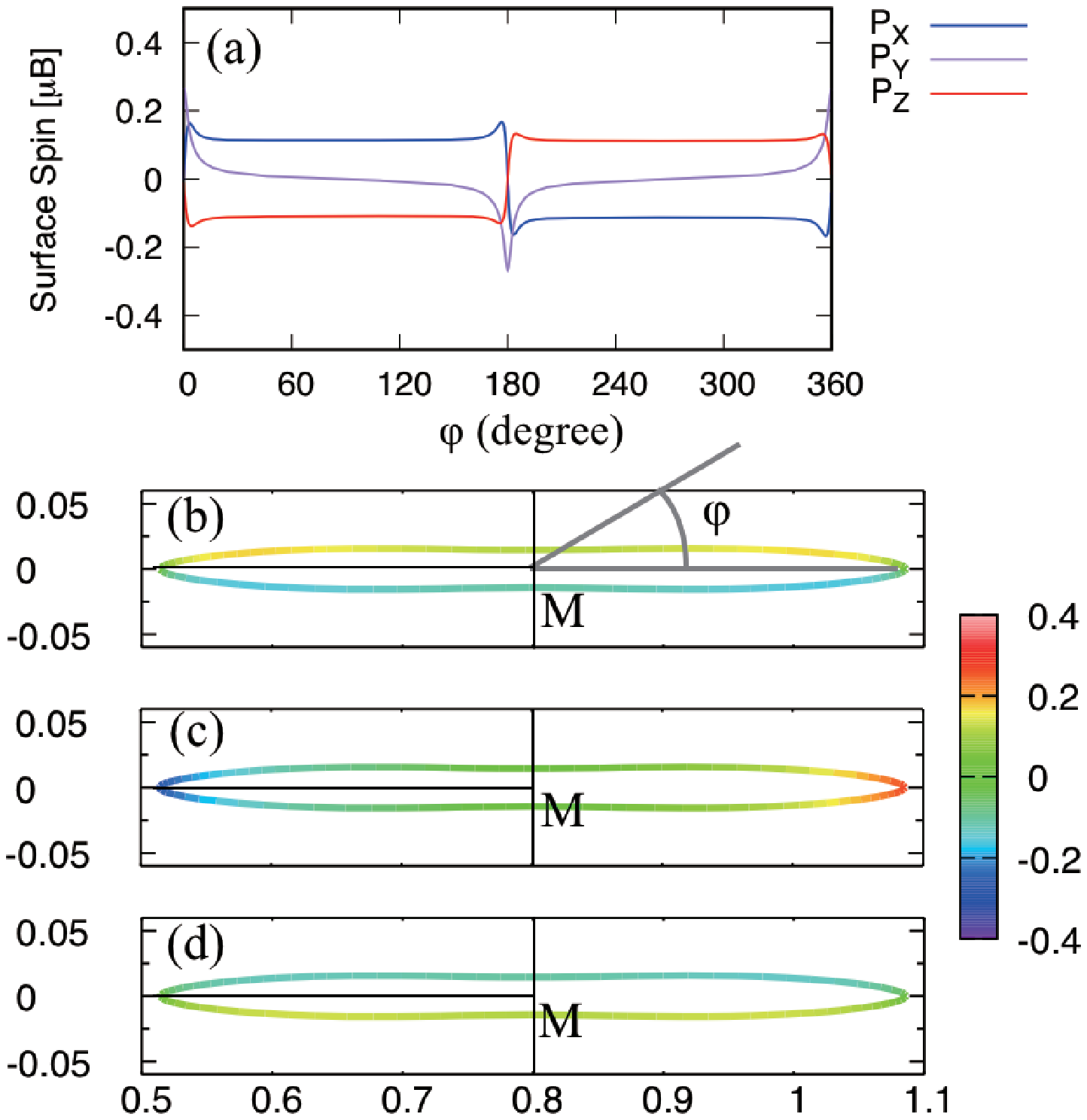}
\end{center}
\caption{Calculated spin texture of S$_3$ pocket. 
(a) Angle-dependet spin components $P_x$, $P_y$, and $P_z$, 
(b) Fermi line with $P_x$ component,
(c) Fermi line with $P_y$ component,
and (d) Fermi line with $P_z$ component.
}
\label{s3}
\end{figure}

\end{document}